\documentclass[epj]{svjour}
\usepackage{graphicx,epsfig, amsmath}
\usepackage{epstopdf}
\begin{document}
\title{Recent breakthroughs in Skyrme-Hartree-Fock-Bogoliubov mass formulas}
\author{S. Goriely\inst{1}, N. Chamel\thanks{Conference presenter}\inst{1}, \and J. M. Pearson\inst{2}  
}
\institute{Institut d'Astronomie et d'Astrophysique, Universit\'e
Libre de Bruxelles, CP226,
1050 Brussels, Belgium 
\and 
D\'eptartement de Physique, Universit\'e de Montr\'eal,
Montr\'eal (Qc) H3C 3J7, Canada
}
\date{Received: date / Revised version: date}
\abstract{
We review our recent achievements in the construction of microscopic mass tables 
based on the Hartree-Fock-Bogoliubov method with Skyrme effective interactions. 
In the latest of our series of HFB-mass models, we have obtained our best fit 
ever to essentially all the available mass data, by treating the pairing more 
realistically than in any of our earlier models. The rms deviation on the 2149 
measured masses of nuclei with $N$ and $Z \ge$ 8 has been reduced for the first 
time in a mean field approach to 0.581 MeV. With the additional constraint on the 
neutron-matter equation of state, this new force is thus very well-suited for 
the study of neutron-rich nuclei and for the description of astrophysical 
environments like supernova cores and neutron-star crusts.
 % end of PACS codes
} %end of abstract
\maketitle
\section{Introduction}
\label{intro}

Nuclear astrophysics applications require the knowledge of 
various nuclear properties (nuclear masses, nuclear level densities (NLD), 
optical potentials, $\gamma$-ray strength functions, etc.) that cannot 
be measured experimentally in the foreseeable future. In order to make 
reliable extrapolations of these quantities far from the domain covered by 
experimental data, we have developed a series of nuclear-mass models based on the 
Hartree-Fock-Bogoliubov (HFB) method with Skyrme and contact-pairing forces, 
together with phenomenological Wigner terms and correction terms for the
spurious collective energy. The model parameters are fitted to essentially 
all the available atomic mass data, requiring that the model reproduce 
several properties of uniform asymmetric nuclear matter determined by 
microscopic calculations with realistic nucleon-nucleon potentials. 
With these nuclear-matter constraints, our models can be reliably applied to 
study astrophysical environments such as supernova cores and neutron star crusts~\cite{lrr}. 
In particular, model HFB-9~\cite{sg05} and all later models constrained 
the underlying Skyrme force to fit the equation of state of neutron matter, 
as calculated by Friedman and Pandharipande~\cite{fp81} for realistic two- 
and three-nucleon forces. 

In this paper, we present our latest models, HFB-16~\cite{cha08} and HFB-17~\cite{gor08}, 
in which we have imposed the additional constraint of reproducing as a function of density 
the $^1S_0$ pairing gaps of uniform asymmetric nuclear matter. This latter constraint
is not only of prime importance for reliable investigations of a possible superfluid
phase in the inner crust of neutron stars, but also turns out to significantly improve
the accurary of the mass fit. 

\section{The Hartree-Fock-Bogoliubov mass formulas}

The Hartree-Fock-Bogoliubov (HFB) mass models HFB-16 and HFB-17 are 
based on the conventional Skyrme force of the form
\begin{eqnarray}
\label{1}
v^{\rm Sky}(\pmb{r_i}, \pmb{r_j})  &=&  t_0(1+x_0 P_\sigma)\delta({\pmb{r}_{ij}}) \nonumber \\
& &+\frac{1}{2} t_1(1+x_1 P_\sigma)\frac{1}{\hbar^2}\left[p_{ij}^2\,\delta({\pmb{r}_{ij}})
+\delta({\pmb{r}_{ij}})\, p_{ij}^2 \right]\nonumber\\
& &+t_2(1+x_2 P_\sigma)\frac{1}{\hbar^2}\pmb{p}_{ij}.\delta(\pmb{r}_{ij})\,
 \pmb{p}_{ij} \nonumber \\
& &+\frac{1}{6}t_3(1+x_3 P_\sigma)\rho(\pmb{r})^\gamma\,\delta(\pmb{r}_{ij})
\nonumber\\
& &+\frac{\rm i}{\hbar^2}W_0(\mbox{\boldmath$\sigma_i+\sigma_j$})\cdot
\pmb{p}_{ij}\times\delta(\pmb{r}_{ij})\,\pmb{p}_{ij}  \quad ,
\end{eqnarray}
where $\pmb{r}_{ij} = \pmb{r}_i - \pmb{r}_j$, $\pmb{r} = (\pmb{r}_i + 
\pmb{r}_j)/2$, $\pmb{p}_{ij} = - {\rm i}\hbar(\pmb{\nabla}_i-\pmb{\nabla}_j)/2$
is the relative momentum, and $P_\sigma$ is the two-body 
spin-exchange operator. The contact pairing force acts only between nucleons of the 
same charge state $q$ ($q = n$ or $p$ for neutron or proton, respectively) and is given by
\begin{equation}
v^{\rm pair}_q(\pmb{r_i}, \pmb{r_j})= v^{\pi\,q}[\rho_n(\pmb{r}),\rho_p(\pmb{r})]~\delta(\pmb{r}_{ij})\quad ,
\label{2}
\end{equation}
where $v^{\pi\,q}[\rho_n,\rho_p]$ is a functional of the nucleon densities.

To the HFB energy calculated for the Skyrme force~(\ref{1}) and the pairing force~(\ref{2})
are added two phenomenological corrections: (i) the Wigner energy~\cite{gtp01,gshpt02}
\begin{eqnarray}
\label{3} 
E_W = V_W\exp\Bigg\{-\lambda\Bigg(\frac{N-Z}{A}\Bigg)^2\Bigg\} \nonumber \\
+V_W^{\prime}|N-Z|\exp\Bigg\{-\Bigg(\frac{A}{A_0}\Bigg)^2\Bigg\} \quad ,
\end{eqnarray}
and (ii) the rotational and vibrational spurious collective energy
\begin{eqnarray}\label{4}
E_{\rm coll}= E_{\rm rot}^{\rm crank}\Big\{b~\tanh(c|\beta_2|) \nonumber  \\ 
+ d|\beta_2|~\exp\{-l(|\beta_2| - \beta_2^0)^2\}\Big\} \quad  ,
\end{eqnarray}
in which $E_{\rm rot}^{\rm crank}$ denotes the cranking-model value of the rotational
correction~\cite{tgpo00} and $\beta_2$ the quadrupole deformation, while all other parameters
are free fitting parameters. The correction term, Eq~(\ref{4}), differs from that 
used in our previous mass models. While the differences are small for large deformations,
the collective energy given by Eq~(\ref{4}) now vanishes for spherical nuclei.

\section{Microscopically derived effective pairing force}

In all of our mass models HFB-1 to HFB-15, the density dependence of the effective 
pairing strength $v^{\pi\,q}[\rho_n,\rho_p]$ acting between nucleons of the same 
charge $q$ ($q=n$ or $p$ for neutrons or protons, respectively) was parametrized 
by an expression of the usual following form
\begin{equation}\label{5}
v^{\pi\,q}[\rho_n,\rho_p] = V_{\pi q}\left\{1-\eta_s
\left(\frac{\rho}{\rho_0}\right)^\alpha\right\}  \quad ,
\end{equation}
where $\rho=\rho_n+\rho_p$ is the total density, $\rho_0$ is the saturation density of 
symmetric nuclear matter, whereas $V_{\pi q}$ is  a free parameter that 
were determined by the global mass fit. The surface parameters , $\eta_s$ and $\alpha$, were 
either set to zero assuming a purely volume pairing force or taken from the work of Ref.~\cite{gar99}. 
Even when constraining the pairing strength $V_{\pi q}$ to a relatively low value in order
 to conform the nuclear pairing with odd-even mass differences, we found that our earlier 
mass models predict unrealistic pairing gaps in uniform nuclear-matter as compared to 
microscopic calculations while yielding comparably good mass fits 
(hence comparably good pairing gaps in finite nuclei). 

In order to improve the treatment of pairing, we have recently  developed a new series 
of HFB mass models by imposing the additional constraint of reproducing as a function of density the $^1S_0$ pairing 
gaps of uniform asymmetric nuclear matter. Fitting all the model parameters
while imposing this constraint would be an extremely onerous numerical task when 
using the ansatz~(\ref{5}). Instead of postulating a density dependence as in 
Eq.~(\ref{5}), we have thus determined the strength of the effective pairing force at 
each neutron and proton density by solving the HFB equations in uniform matter 
and requiring that the resulting gap reproduce exactly the microscopic pairing 
gap calculated with realistic forces at that neutron and proton density. 
In this way, the pairing strength for the nucleon species $q$ is given by
\begin{eqnarray}
\label{6}
v^{\pi\,q}[\rho_n, \rho_p]=-8\pi^2\left(\frac{\hbar^2}
{2 M_q^*(\rho_n, \rho_p)}\right)^{3/2} \times \nonumber \\
\times \left(\int_0^{\mu_q+\varepsilon_{\Lambda}}{\rm d}\xi
\frac{\sqrt{\xi}}{\sqrt{(\xi-\mu_q)^2+\Delta_q(\rho_n, \rho_p)^2}}
\right)^{-1}  \quad ,
\end{eqnarray}
where $\Delta_q(\rho_n,\rho_p)$ is the corresponding pairing gap in uniform matter, 
$M_q^*(\rho_n, \rho_p)$ is the effective nucleon mass and $\varepsilon_{\Lambda}$ is 
the pairing cutoff. The chemical potential $\mu_q$ is approximated by 
\begin{equation}
\label{7}
\mu_q = \frac{\hbar^2 k_{{\rm F}q}^2}{2 M_q^*} \, ,
\end{equation}
where $k_{{\rm F}q} = (3\pi^2 \rho_q)^{1/3} $ is the nucleon Fermi wave number.

Given the discrepancies between the various microscopic many-body calculations of 
the $^1S_0$ pairing gaps in uniform nuclear matter, we have considered two different 
cases. In model HFB-16, we have taken the microscopic pairing gap calculated at the 
lowest BCS level using the Argonne v14 potential~\cite{lom01} and following Duguet~\cite{dg04}, 
we have assumed that the pairing force for nucleons $q$ depends only on the density $\rho_q$. We have dropped 
this assumption in our latest model HFB-17 and we have considered the recent Brueckner calculations of 
Ref.~\cite{cao06} which include the effect on the interaction of medium polarization. 
These calculations were performed using the Argonne v18 potential both with and without self-energy corrections. 
Adopting a microscopic pairing gap calculated with self-energy corrections is quite challenging since ideally for 
consistency the Skyrme effective mass $M_q^*$ should be fitted to the corresponding microscopic 
effective mass. However this seems to be impossible within the framework of the conventional 
Skyrme forces used here. For this reason, we have used the microscopic pairing gap of Ref~\cite{cao06} 
including only medium-polarization effects and we have set $M_q^* = M$ in Eqs.~(\ref{6}) and (\ref{7}). 

\begin{figure}
% Use the relevant command for your figure-insertion program
% to insert the figure file.
% For example, with the option graphics use
\resizebox{0.4\textwidth}{!}{%
  \includegraphics{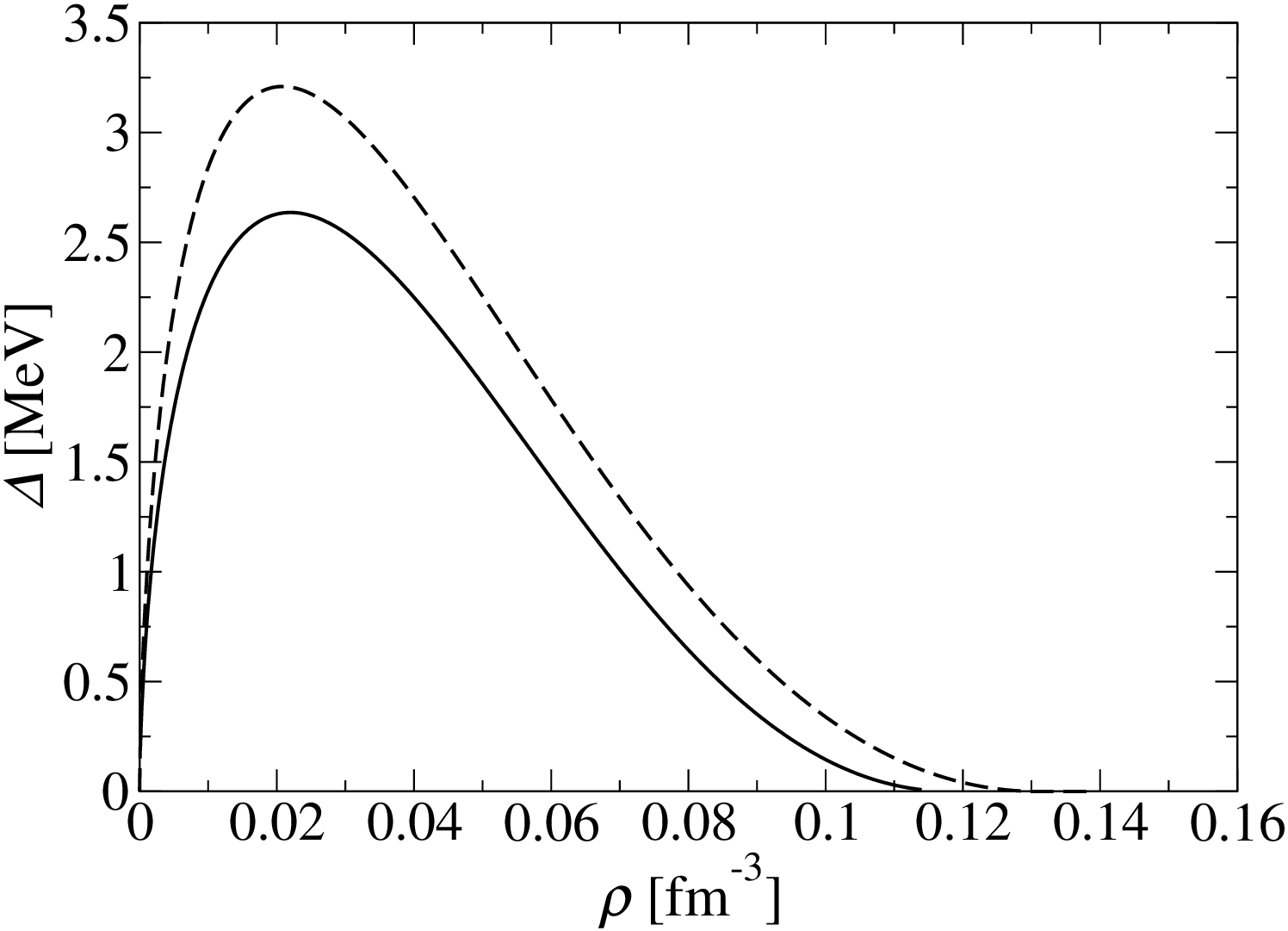}
}
% If not, use
%\vspace{5cm}       % Give the correct figure height in cm
\caption{$^1S_0$ neutron pairing gap $\Delta$ in infinite neutron matter
as a function of the density $\rho$. The dashed line represents the gap 
obtained in the BCS approximation from Ref.~\cite{lom01} and used in model 
HFB-16, while the solid line is the gap calculated in Ref.~\cite{cao06} 
including the effect on the interaction of medium polarization and used in model 
HFB-17.}
\label{fig:1}       % Give a unique label
\end{figure}

\begin{figure}
% Use the relevant command for your figure-insertion program
% to insert the figure file.
% For example, with the option graphics use
\resizebox{0.4\textwidth}{!}{%
  \includegraphics{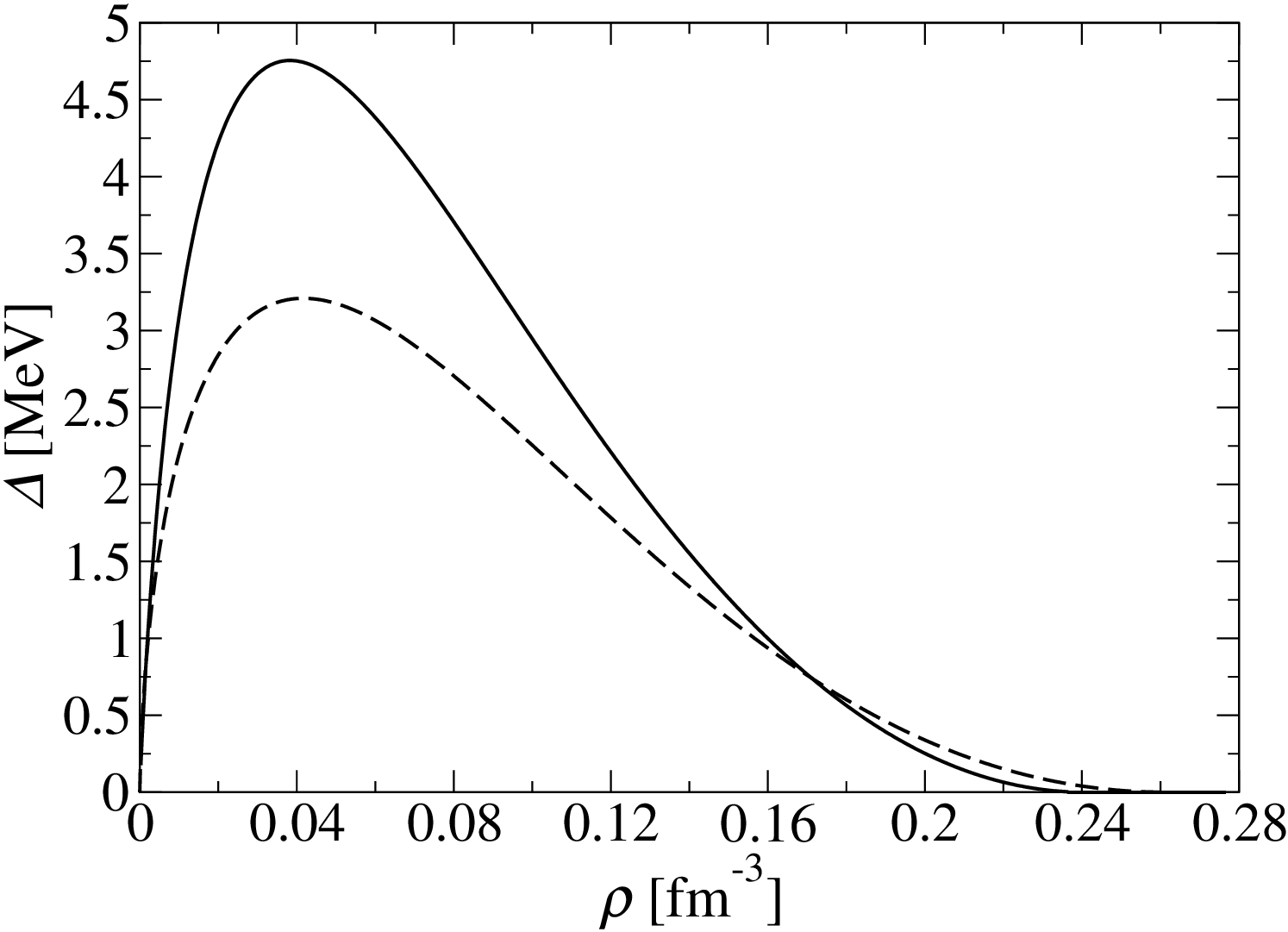}
}
% If not, use
%\vspace{5cm}       % Give the correct figure height in cm
\caption{$^1S_0$ neutron pairing gap $\Delta$ in infinite symmetric nuclear matter
as a function of the density $\rho$. The dashed line represents the gap 
obtained in the BCS approximation from Ref.~\cite{lom01} and used in model 
HFB-16, while the solid line is the gap calculated in Ref.~\cite{cao06} 
including the effect on the interaction of medium polarization and used in model HFB-17.}
\label{fig:2}       % Give a unique label
\end{figure}

Ref.~\cite{cao06} calculates pairing gaps only for symmetric nuclear matter, 
$\Delta_{SM}(\rho = \rho_n+\rho_p)$, and pure neutron matter, 
$\Delta_{NM}(\rho_n)$. Since we need the pairing gaps for arbitrary asymmetry
we adopted the interpolation ansatz
\begin{equation}
\Delta_q(\rho_n,\rho_p)=\Delta_{SM}(\rho)(1-|\eta|) \pm \Delta_{NM}(\rho_q)
\,\eta\,\frac{\rho_q}{\rho}   \, ,
\end{equation}
where $\eta = (\rho_n-\rho_p)/\rho$ and the upper (lower) sign is to be taken 
for $q = n (p)$; we have also assumed charge symmetry, i.e., 
$\Delta_n(\rho_n,\rho_p) = \Delta_p(\rho_p,\rho_n)$. This expression ensures 
that for symmetric nuclear matter, $\Delta_q(\rho/2,\rho/2)=\Delta_{SM}(\rho)$ and for 
neutron matter $\Delta_n(\rho, 0)=\Delta_{NM}(\rho)$ and $\Delta_p(\rho, 0)=0$.
For convenience, we have used the essentially exact analytical representations
\begin{equation}
\Delta_{\rm SM}(\rho)=\theta(k_{\rm m}-k_{{\rm F}})\, \Delta_0 \frac{k_{{\rm F}}^3}{k_{{\rm F}}^2+k_1^2}\frac{(k_{{\rm F}}-k_2)^2}{(k_{{\rm F}}-k_2)^2+k_3^2}\, ,
\end{equation}
\begin{equation}
\Delta_{\rm NM}(\rho_n)=\theta(k_{\rm m}-k_{{\rm F}n})\, \Delta_0 \frac{k_{{\rm F}n}^2}{k_{{\rm F}n}^2+k_1^2}\frac{(k_{{\rm F}n}-k_2)^2}{(k_{{\rm F}n}-k_2)^2+k_3^2}\, ,
\end{equation}
where $k_{\rm F}=(3\pi^2 \rho/2)^{1/3}$, $\theta$ is the Heaviside unit-step function, and the associated parameters are given 
in Table~\ref{tab1}. 
The different pairing gaps are shown in Figs.~\ref{fig:1} and \ref{fig:2}. 
For odd-nucleon nuclei we make use of the ``equal-filling'' approximation~\cite{mr08}. 
In order to take phenomenologically into account the contribution to pairing of the 
time-odd fields~\cite{ber08}, we multiply the pairing strength by a parity factor 
$f_q^{\pm}$ ($+$ for an even number of nucleons, $-$ otherwise) that we allow to be 
different for neutrons and protons due to Coulomb forces and possible charge symmetry 
breaking effects. By definition we set $f_n^+=1$. 

\begin{figure}
% Use the relevant command for your figure-insertion program
% to insert the figure file.
% For example, with the option graphics use
\resizebox{0.4\textwidth}{!}{%
  \includegraphics{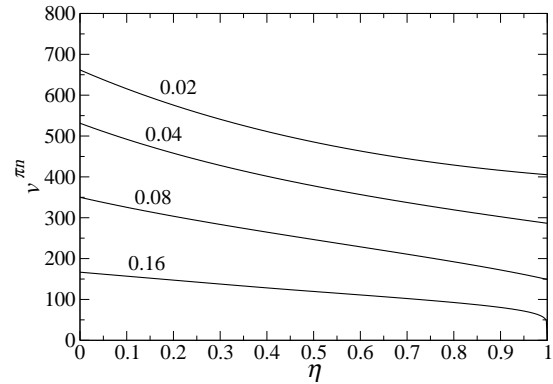}
}
% If not, use
%\vspace{5cm}       % Give the correct figure height in cm
\caption{Isospin dependence of the effective neutron pairing strength 
$v^{\pi n}$ of model HFB-17 for different densities $\rho$ (indicated above each curve 
in fm$^{-3}$).}
\label{fig:3}       % Give a unique label
\end{figure}

\begin{figure}
% Use the relevant command for your figure-insertion program
% to insert the figure file.
% For example, with the option graphics use
\resizebox{0.4\textwidth}{!}{%
  \includegraphics{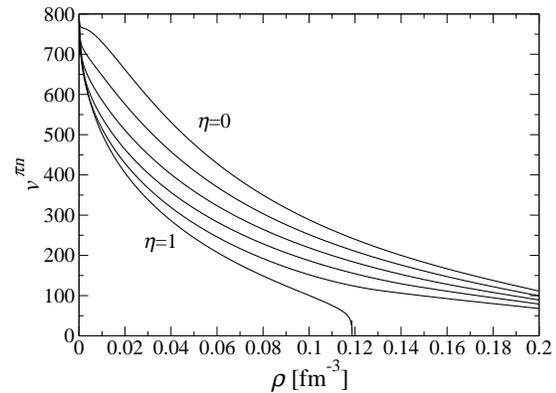}
}
% If not, use
%\vspace{5cm}       % Give the correct figure height in cm
\caption{Effective neutron pairing strength $v^{\pi n}$ of model HFB-17 
as a function of density $\rho$ for different isospin asymmetries $\eta$ 
by step of 0.2.}
\label{fig:4}       % Give a unique label
\end{figure}

The corresponding effective pairing force, as given by Eq.~(\ref{6}), is 
shown in Figs.~\ref{fig:3} and \ref{fig:4} for the 
HFB-17 mass model. 

\begin{table}
\centering
\caption{Parameters of the analytical fit for the microscopic reference 
pairing gaps (the unit of length is fermi and the unit of energy is MeV).}
\label{tab1} 
\vspace{.5cm}
\begin{tabular}{|c|c|c|c|c|c|}
\hline & $\Delta_0$ & $k_1$ & $k_2$ & $k_3$ & $k_{\rm m}$ \\ 
% \hline BCS & 910.603 & 1.38297 & 1.57068 & 0.905237 & 1.57 \\ 
\hline SM & 133.779 & 0.943146 & 1.52786 & 2.11577 & 1.51 \\ 
\hline NM & 14.9003 & 1.18847 & 1.51854 & 0.639489 & 1.52 \\ \hline 
\end{tabular}
\end{table}

% For one-column wide figures use

\section{Results of the mass fit}
\begin{figure}
\centering
\resizebox{0.95\columnwidth}{!}{ \includegraphics{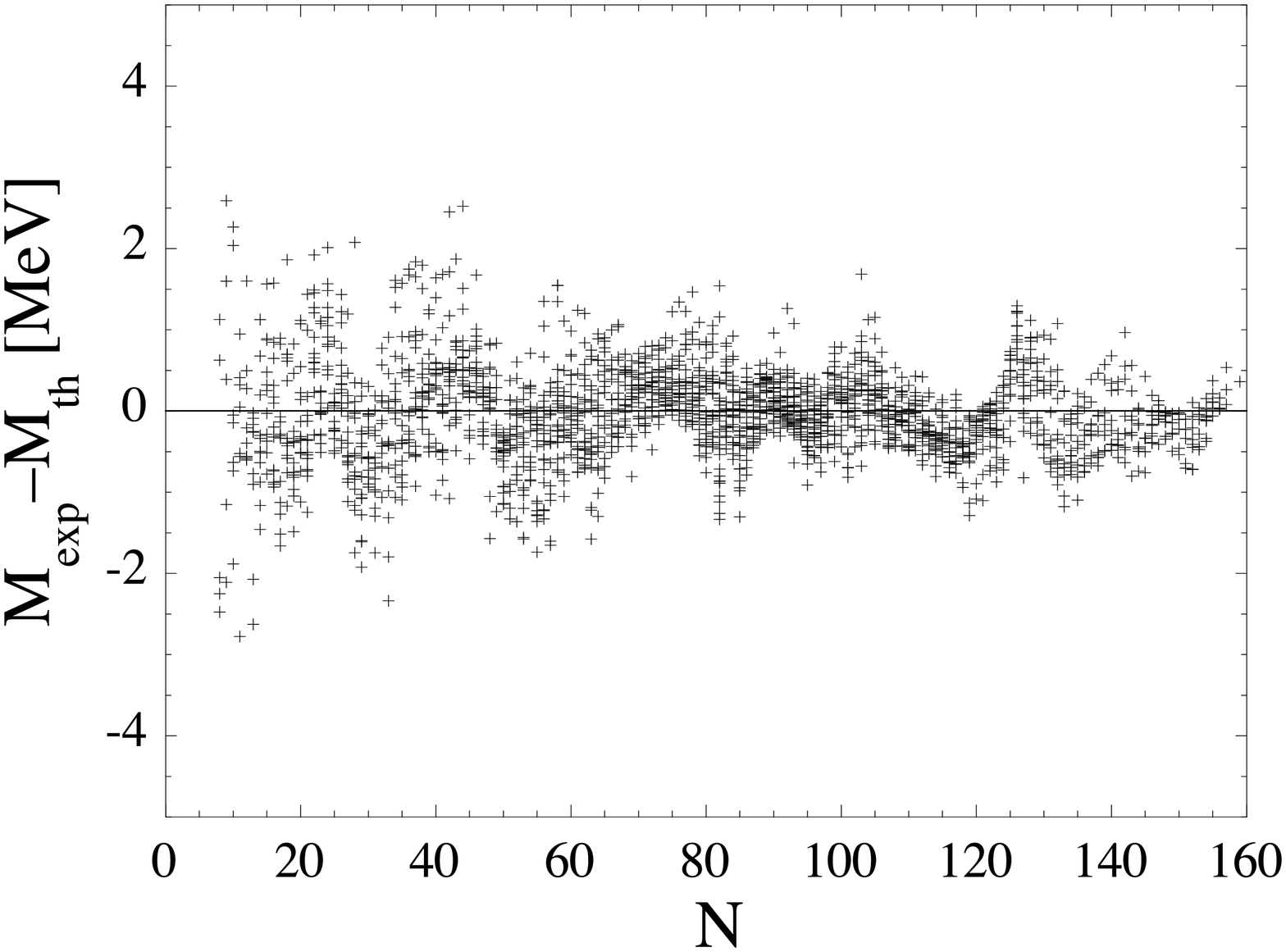}}
\caption{Differences between experimental and calculated masses as a function 
of the neutron number $N$ for the HFB-17 mass model. }
\label{fig:5}
\end{figure}

As explained in the previous sections, the HFB-17 mass model~\cite{gor08} represents our latest attempt 
to provide a universal force capable of predicting globally static nuclear structure properties. 
The 10 parameters of the Skyrme force along with the 4 parameters of the pairing force (including the 
pairing cutoff) and the 9 parameters of the Wigner and collective corrections have been fitted on 
the 2149 measured masses of nuclei with $N$ and $Z \ge$ 8 given in the 2003 AME~\cite{audi03}. 
The deviations between the experimental data and the HFB-17 predictions are shown graphically in 
fig.~\ref{fig:5}. The rms and mean (data - theory) values of these deviations are  0.581~MeV and -0.019~MeV, 
respectively. HFB-17 is not only the most accurate  mass model ever achieved within the mean-field framework, 
it also satisfies extra physical constraints that make it more suitable for astrophysics applications. 
In particular, the calculated quadrupole  moments, charge radii, charge-density distributions and 
spins are also found to be in good  agreement with experiment. As done in our previous mass fits, 
when determining the Skyrme force parameters, a special attention has been paid to properly describe 
the properties (not only the pairing) of infinite nuclear and neutron matter determined from realistic 
calculations~\cite{fp81,lom08,zuo06}. The excellent fit of HFB-17 to the neutron matter curve of 
Ref.~\cite{fp81} was achieved simply by imposing a nuclear-matter symmetry coefficient of $J$ = 30 MeV. 
The potential energy per particle determined with the BSk17 Skyrme force was also found to be in fair 
agreement with the recent calculations of Ref.~\cite{lom08} in each of the four two-body spin-isospin 
$(S,T)$ channels and for all densities. Finally, in contrast to most of the traditional Skyrme forces, 
our BSk17 (and previous forces) is consistent with the isovector splitting of the effective mass 
deduced from measurements of isovector giant resonances and confirmed in several many-body calculations 
with realistic forces~\cite{zuo06}. More details on these comparisons can be found in Ref.~\cite{gor08}.

\begin{figure}
\centering
\resizebox{0.95\columnwidth}{!}{ \includegraphics{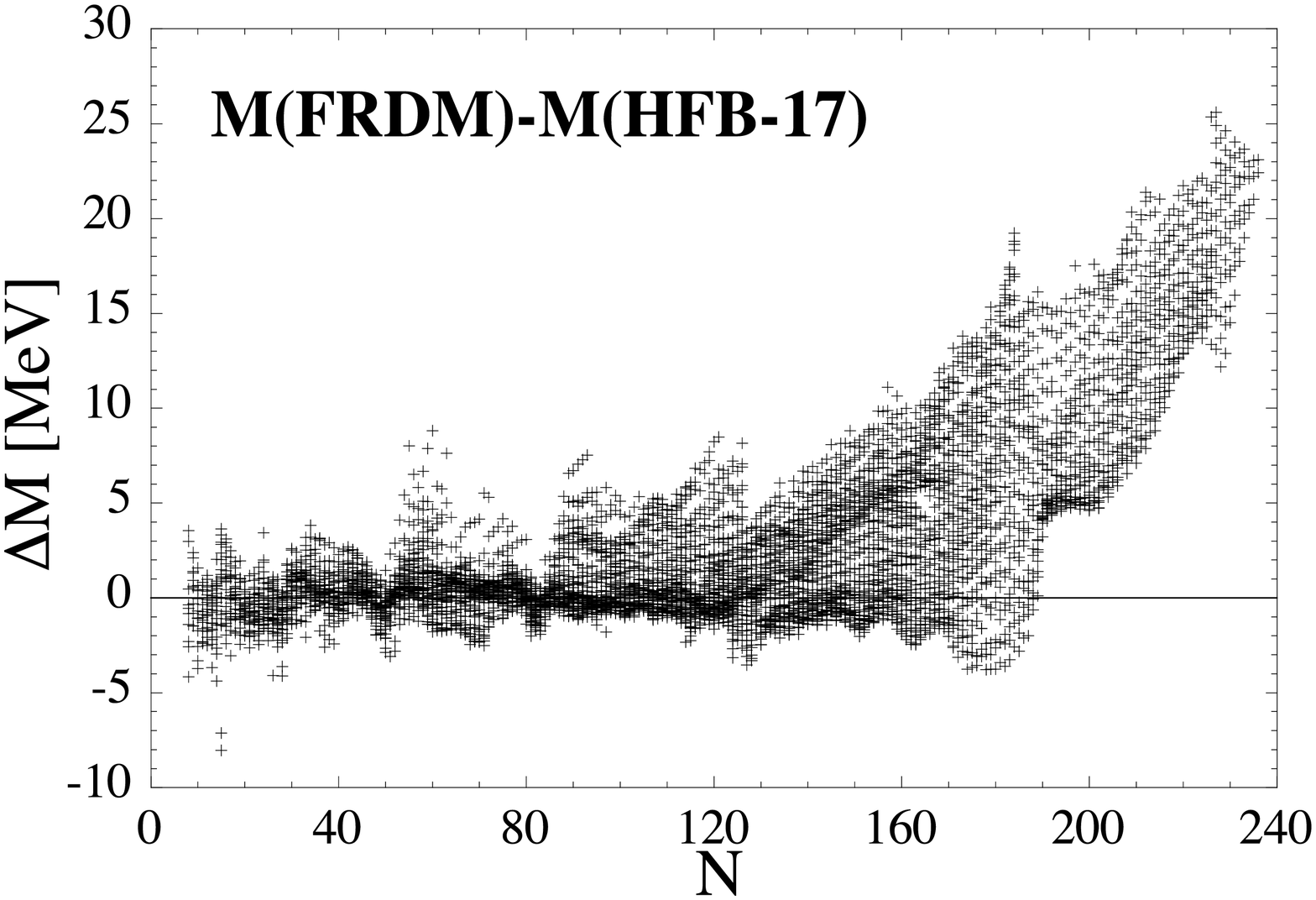}}
\caption{Differences between the FRDM and HFB-17 mass predictions as a function
of (left) $N$   for all $8 \le Z \le 110$ nuclei lying between the proton and neutron drip lines.}
\label{fig:6}
\end{figure}

\begin{figure}
\centering
\resizebox{0.95\columnwidth}{!}{ \includegraphics{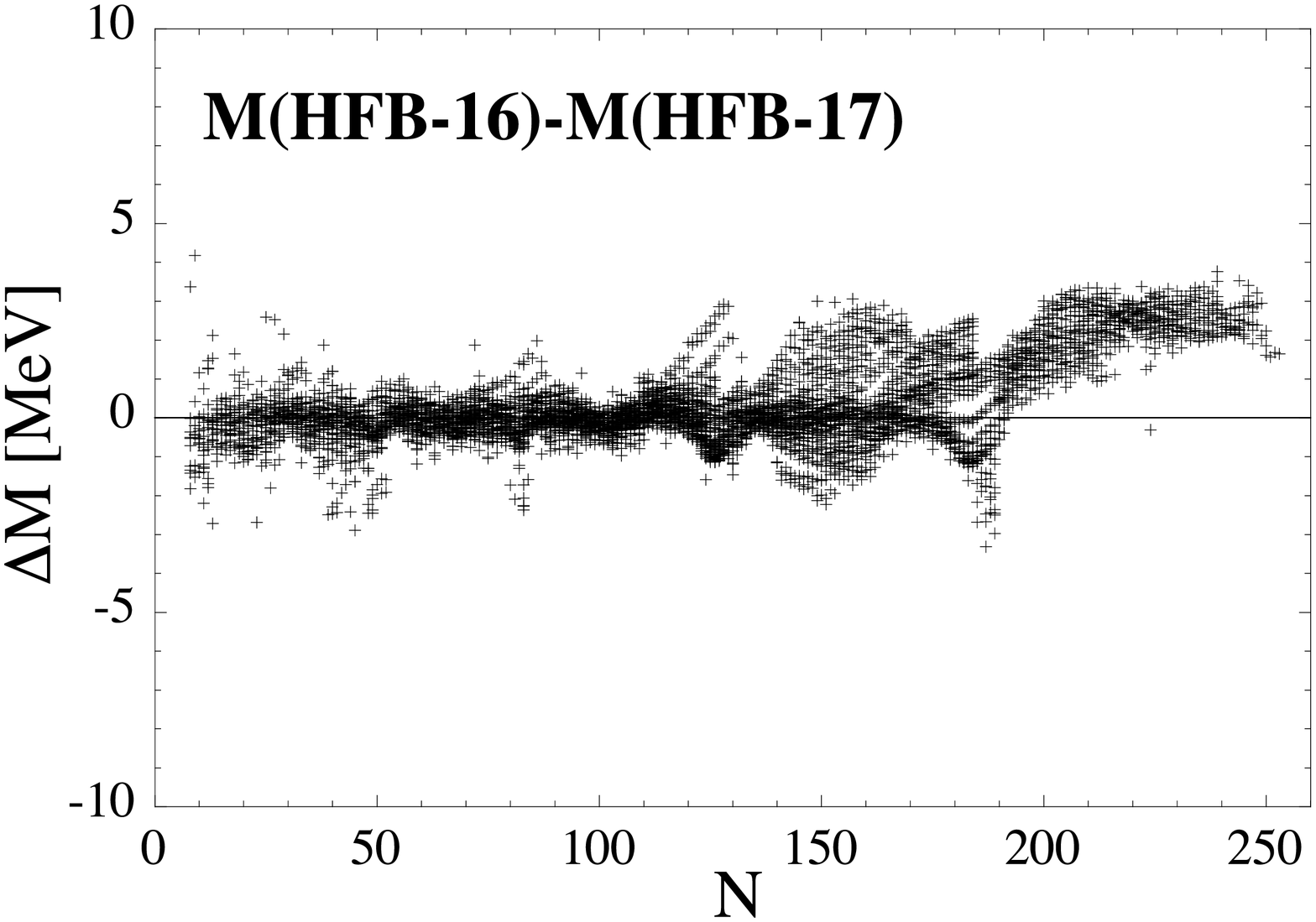}}
\caption{Differences between the HFB-16 and HFB-17 mass predictions as a function $N$  for all $8 \le Z \le 110$ nuclei lying between the proton and neutron drip lines.}
\label{fig:7}
\end{figure}

In Fig. \ref{fig:6} we compare the HFB-17 predictions with those of the finite range droplet model 
(FRDM)~\cite{frdm}  for all $8 \le Z \le 110$ nuclei lying between the proton and neutron  drip lines. 
Significant differences (up to 25~MeV) are found, especially for heavy  neutron-rich nuclei. 
Such differences are much larger than what is obtained between our various HFB mass models, 
as illustrated in Fig.~\ref{fig:7}. 

\begin{figure}
\centering
\resizebox{0.95\columnwidth}{!}{ \includegraphics{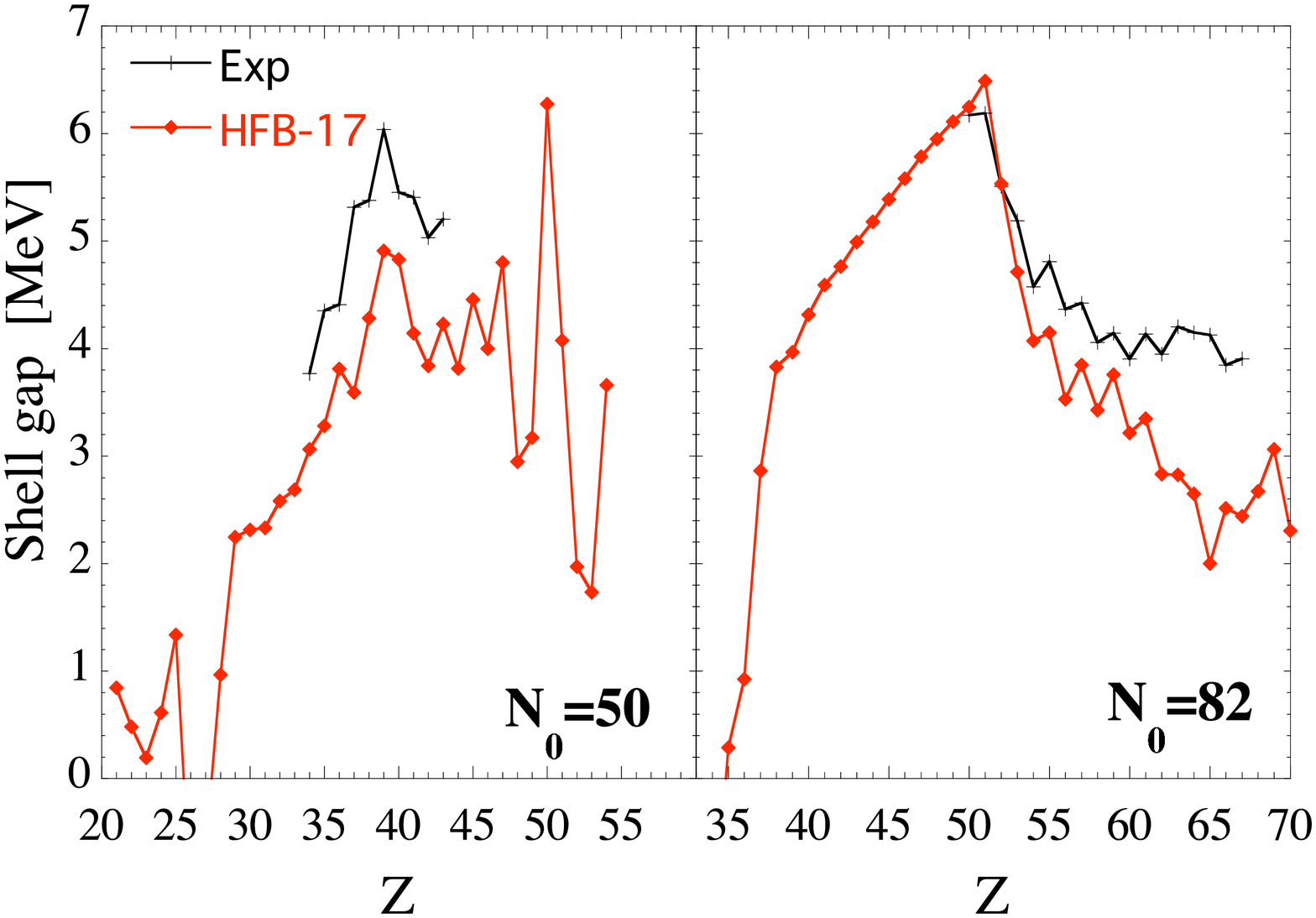}}
\caption{$N_0$ = 50 and $N_0=82$ shell gaps as function of $Z$ for the mass models HFB-17.}
\label{fig:8}
\end{figure}

\begin{figure}
\centering
\resizebox{0.95\columnwidth}{!}{ \includegraphics{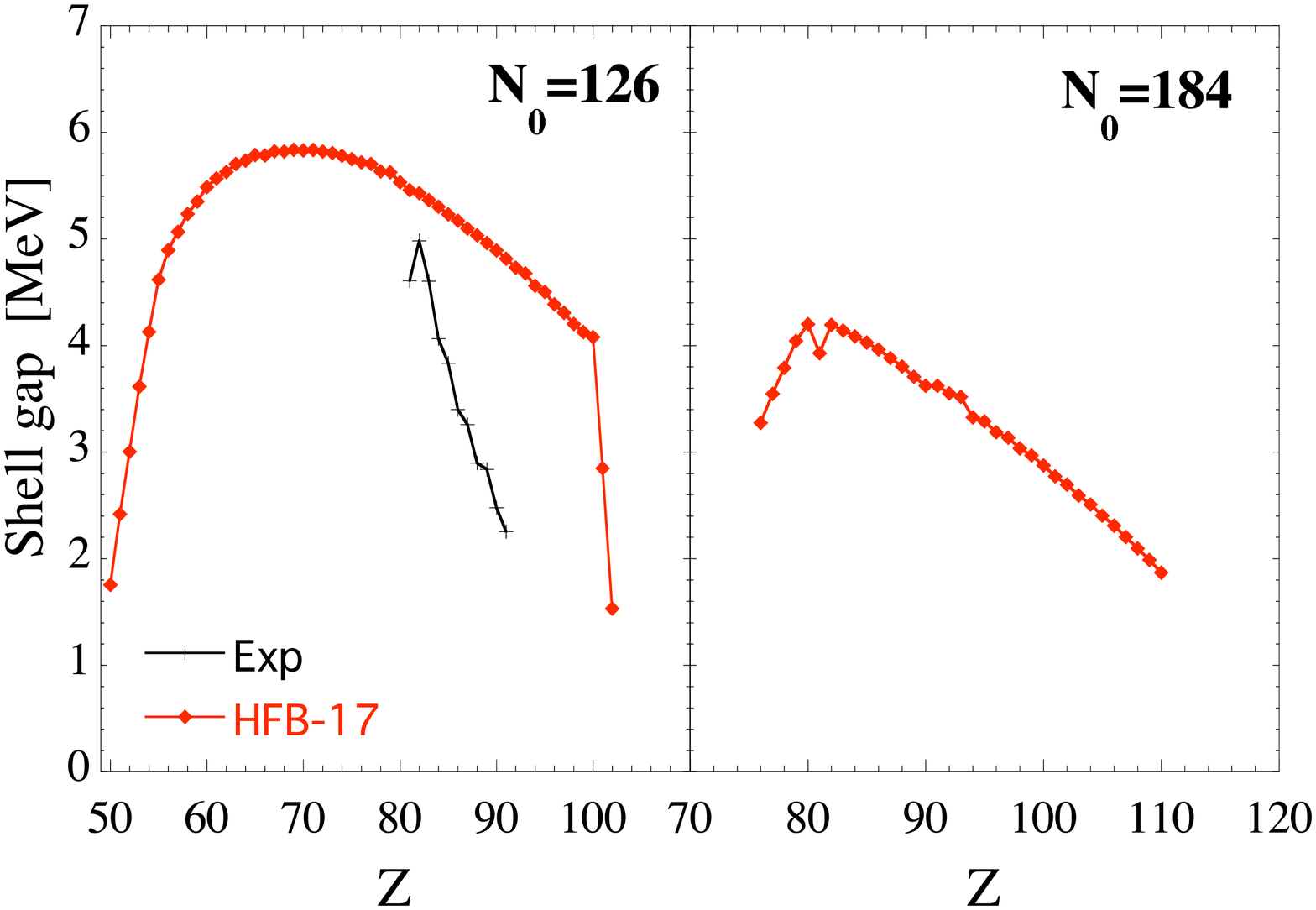}}
\caption{$N_0$ = 126 and $N_0=184$ shell gaps as function of $Z$ for the mass models HFB-17.}
\label{fig:9}
\end{figure}

\begin{figure}
\centering
\resizebox{0.95\columnwidth}{!}{ \includegraphics{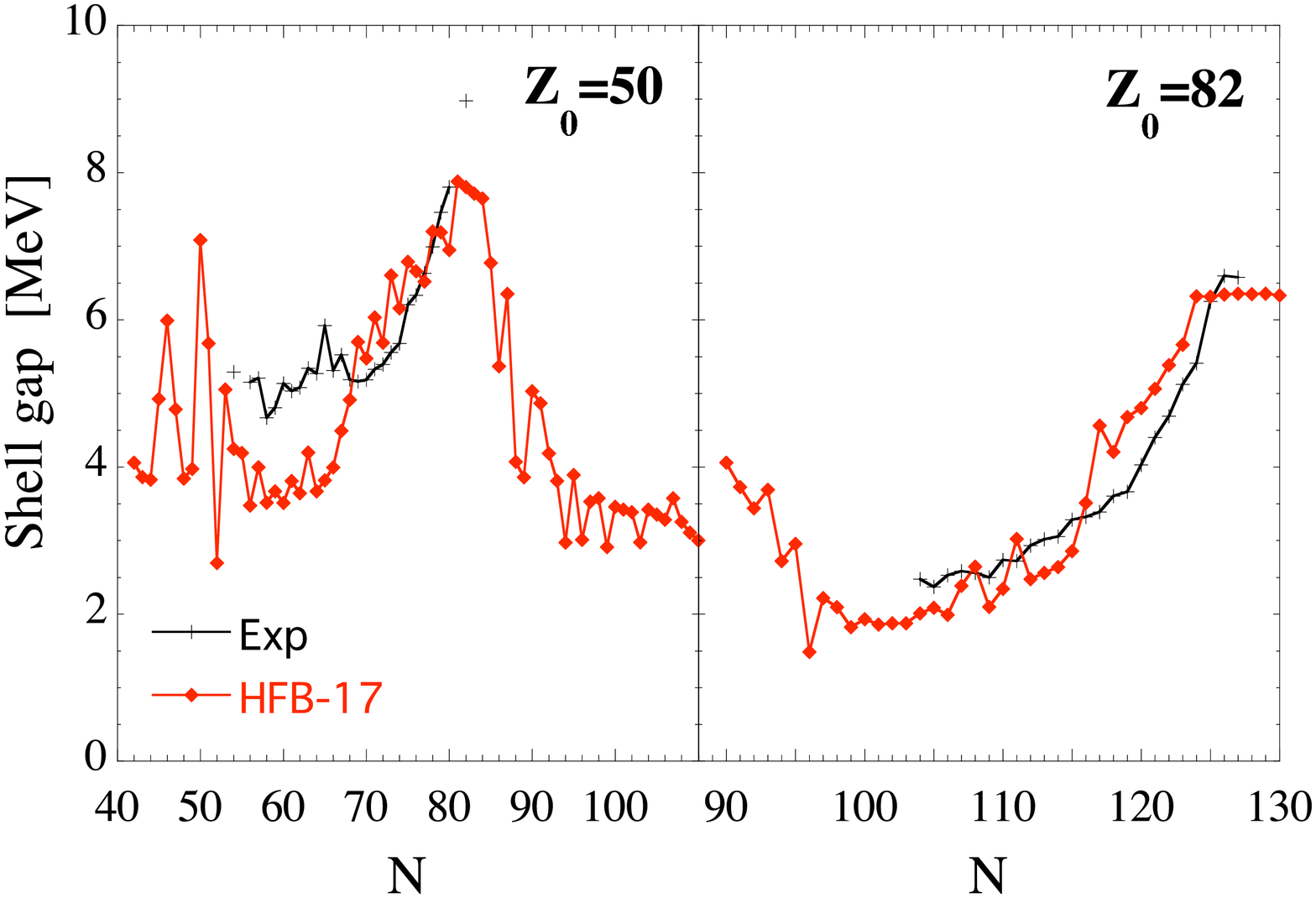}}
\caption{$Z_0$ = 50 and $Z_0=82$ shell gaps as function of $N$ for the mass models HFB-17.}
\label{fig:10}
\end{figure}

In Figs. \ref{fig:8} and \ref{fig:9}, we show the neutron-shell gaps, defined by
\begin{equation}
\label{shell}
\Delta_n(N_0, Z) = S_{2n}(N_0,Z) - S_{2n}(N_0+2,Z)  \quad ,
\end{equation}
 as a function of $Z$ for the magic numbers $N_0 = 50, 82, 126$ and $184$ for the HFB-17 model 
($S_{2n}$ is the two-neutron separation energy). Likewise in Fig.~\ref{fig:10}  we show the 
proton-shell gaps as a function of $N$ for $Z_0$ = 50 and 82. Here we consider just the comparison 
with the data. Given the overall close agreement between experimental and theoretical shell gaps, the differences 
found for the $N_0=126$ neutron-shell gaps clearly emerge as unsatisfactory, especially in the vicinity 
of doubly-magic $^{208}$Pb nuclei. It is even more disturbing that the $Z_0=82$ proton-shell gaps agree 
rather well. The double magicity around $^{208}$Pb remains unresolved in all mean field calculations, 
regardless of the type of force or its parametrization.

\section{Astrophysical applications}

\begin{figure}
% Use the relevant command for your figure-insertion program
% to insert the figure file.
% For example, with the option graphics use
\resizebox{0.4\textwidth}{!}{%
  \includegraphics{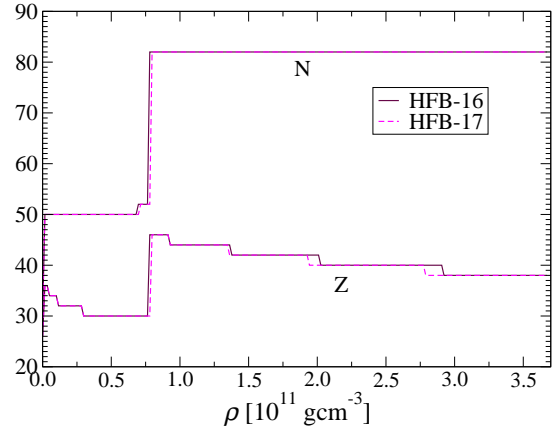}
}
% If not, use
%\vspace{5cm}       % Give the correct figure height in cm
\caption{Proton number $Z$ and neutron number $N$ of nuclei in the outer crust of neutron stars 
for the HFB-16 and HFB-17 mass models.}
\label{fig:11}       % Give a unique label
\end{figure}

Following the classical paper of Baym, Pethick and Sutherland~\cite{bps71}, we have calculated 
the structure of the outer crust of neutron stars for the HFB-16 and HFB-17 mass models. 
The only microscopic inputs are the values of the atomic masses. We have used 
experimental data when available. The results are shown in Fig.~\ref{fig:11}. The composition of 
the nuclei in the outer crust is essentially the same for the two mass models HFB-16 and HFB-17. 
Both models predict that at densities above $\rho_{\rm ND} \simeq 3.7\times 10^{11} $ gcm$^{-3}$, 
neutrons start to drip out of nuclei. The corresponding layer of the outer crust is composed of 
nuclei with $Z=38$ and $N=82$. 

\begin{figure}
% Use the relevant command for your figure-insertion program
% to insert the figure file.
% For example, with the option graphics use
\resizebox{0.4\textwidth}{!}{%
  \includegraphics{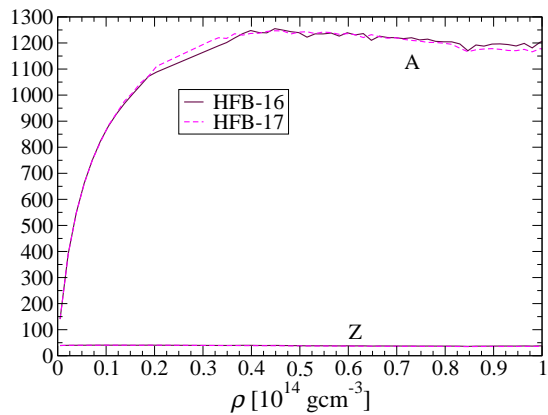}
}
% If not, use
%\vspace{5cm}       % Give the correct figure height in cm
\caption{Proton number $Z$ and mass number $A$ of ``clusters'' in the inner crust of neutron stars 
for the HFB-16 and HFB-17 mass models. Note that $A$ includes not only the number of 
bound nucleons but also the mean number of free neutrons per lattice site.}
\label{fig:12}       % Give a unique label
\end{figure}

\begin{figure}
% Use the relevant command for your figure-insertion program
% to insert the figure file.
% For example, with the option graphics use
\resizebox{0.4\textwidth}{!}{%
  \includegraphics{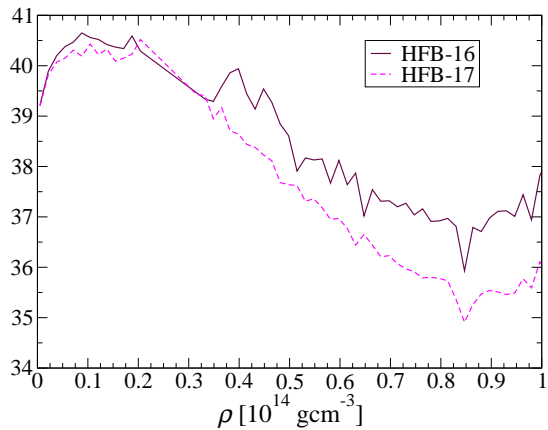}
}
% If not, use
%\vspace{5cm}       % Give the correct figure height in cm
\caption{Proton number $Z$ of nuclei in the inner crust of neutron stars 
for the HFB-16 and HFB-17 mass models.}
\label{fig:13}       % Give a unique label
\end{figure}

The structure of the inner crust of neutron stars, at densities above $\rho_{\rm ND}$, is not simply
determined by atomic masses due to the presence of the neutron gas. The free neutrons modify the properties 
of the ``nuclei'' by i) exerting a pressure on them and ii) by reducing their surface tension~\cite{lrr}. 
In order to make reliable predictions of the composition of the neutron star crust beyond neutron drip, both 
the nucleons bound inside clusters and the free neutrons have to be described consistently. For this
purpose, we have applied the Extended Thomas-Fermi method up to the 4th order (see Ref.~\cite{onsi08} for details) 
with the Skyrme effective forces BSk16 and BSk17 underlying the HFB-16 and HFB-17 mass models respectively.
As can be seen in Fig.~\ref{fig:12}, the two models yield very similar results. The properties of the 
neutron star crust (elastic constants, electric and thermal conductivities, etc.) are strongly dependent 
on the proton number $Z$ of the nuclear clusters. As shown in Fig.~\ref{fig:13}, the differences in 
$Z$ between the two crust models do not exceed $2$ units.

\section{Conclusions}

Despite all the restrictions imposed on the model parameter space, we have 
obtained the most accurate mass table ever achieved within the mean field framework, 
the rms deviation falling to 0.581 MeV. Given also the constraint imposed on the Skyrme force by
microscopic calculations of nuclear and neutron matter, this new model is particularly 
well adapted to astrophysical applications involving a neutron-rich environment,
such as the elucidation of the r-process of nucleosynthesis, and the
description of supernova cores and neutron-star crusts. The HFB-17 mass table is 
made available to the scientific community at http://www-astro.ulb.ac.be.

\end{document}